\documentstyle[12pt,a4wide,epsf]{article}
\begin{document}

\begin{center}
{\Large\bf Warm Quintessential Inflation}

\bigskip

{\large Konstantinos Dimopoulos}\footnote{\tt k.dimopoulos1@lancaster.ac.uk}
{\large and Leonora Donaldson-Wood}\footnote{\tt
  l.donaldsonwood@lancaster.ac.uk}

\medskip

{\sl Consortium for Fundamental Physics}

{\sl Physics Department, Lancaster University}

{\sl Lancaster, LA1 4YB, UK}

\end{center}

\begin{abstract}
  We introduce warm quintessential inflation and study it in the weak
  dissipative regime. We consider the original quintessential inflation model,
  which approximates quartic chaotic inflation at early times and thawing
  quartic inverse-power-law quintessence at present. We find that the model
  successfully accounts for both inflation and dark energy observations, while
  it naturally reheats the Universe, thereby overcoming a major problem of
  quintessential inflation model-building.
\end{abstract}

\section{Introduction}

Inflation is overwhelmingly the best mechanism for explaining the observed
structure in the Universe as well as its spatial flatness and large-scale
homogeneity \cite{inflation}. In the same time, the discovery of dark energy
\cite{DEobs} is best attributed to a non-zero, albeit incredibly fine-tuned,
cosmological constant in the benchmark paradigm of $\Lambda$CDM \cite{LCDM}.
However, recently both proposals have been challenged by the swampland
conjectures \cite{swampland}, which stipulate the impossibility of de-Sitter
vacua in string theory and also set stringent constraints on inflation
model-building and undermine $\Lambda$CDM \cite{cosmoswamp} (but see also
Ref.~\cite{lindeswamp}). Such constraints
are not possible to meet with conventional inflation \cite{infswamp}. A
successful way to model inflation while satisfying the swampland conjectures is
incorporating dissipating effects \cite{warmswamp}, as in warm inflation
\cite{warm}. On the dark energy front, the observations of the current
accelerated expansion can be
explained by quintessence instead of a non-zero cosmological constant $\Lambda$
\cite{quint}, which is also in agreement with the swampland conjectures
\cite{qswamp}. In this letter, we attempt to join the two and introduce warm
quintessential inflation 
(for a reference list on quintessential inflation see
Refs.~\cite{QIalpha,samietal}),
which has the additional advantage of providing a natural mechanism
for reheating the Universe. Reheating is of particular significance in
quintessential inflation because the conventional reheating
by the decay of the inflaton field at the end of inflation cannot occur as the
field needs to survive until the present and become quintessence.
We use natural units where \mbox{$c=\hbar=k_{\rm B}=1$} and
\mbox{$8\pi G=m_P^{-2}$}, where \mbox{$m_P=2.43\times 10^{18}\,$GeV} is the
reduced Planck mass.

\section{The model}

The original quintessential inflation model is \cite{PV}\footnote{There is
  recently revamped interest in this model, see for example Ref.~\cite{recent}.}
\begin{equation}
  V(\phi)=
  \left\{
  \begin{array}{ll}
\lambda(\phi^4+M^4) & {\rm for}\quad\phi<0\\
\frac{\mbox{\normalsize $\lambda M^8$}}{\mbox{\normalsize $\phi^4+M^4$}} &
     {\rm for}\quad\phi>0
\end{array}\right.\,,
\label{V}
\end{equation}
where \mbox{$0<M\ll m_P$}. For negative values of the inflaton field
\mbox{$\phi\ll -M$}, the above potential reduces to quartic chaotic inflation,
which has been excluded by observations unless it is ``warmed up'', by
considering significant dissipation effects. During inflation
\mbox{$\phi\sim -m_P$}. For positive values of the field \mbox{$\phi\gg M$}
the potential becomes inverse power-law (IPL) quintessence. Such quintessence
models feature a tracker solution, which however, is too steep to satisfy
observations in the case of an inverse quartic potential
\mbox{$V\propto\phi^{-4}$}. However, in our case, the field does not follow the
tracker but, after the end of inflation, it rushes down its runaway potential
and freezes at a value \mbox{$\phi_F\sim m_P$} with some residual potential
density, which explains dark energy. At present, the field unfreezes and begins
slowly rolling down its potential. Such quintessence is called ``thawing''
\cite{thawing}.

While the field runs from inflation at \mbox{$\phi\sim -m_P$} to quintessence at
\mbox{$\phi\sim m_P$} it is kinetically dominated and oblivious of the
potential \cite{kination}. Thus, the awkward discontinuity (in the fourth
derivative) of the potential in Eq.~(\ref{V}) is not felt.
In fact, the potential in Eq.~(\ref{V}) is only experienced by the field when
\mbox{$|\phi|\sim m_P$}, which means that Eq.~(\ref{V}) is only a guideline to
the actual form of $V(\phi)$ and should not be taken too seriously.

In addition, the field runs over super-Planckian distance from the end of
inflation to its eventual freezing. It is likely that the dissipative properties
of the field are different in these two different patches of the scalar
potential, which are several Planck scales apart. Indeed, we assume that
dissipative effects are important only when the field is slow-rolling during 
inflation with \mbox{$\phi\sim -m_P$}. Additionally, we consider only the
weak dissipative regime, where the dynamics of the field are not affected by
dissipation (no extra friction) so this issue is not of our concern.

In the weak dissipative regime, the only effect of dissipation is that the
quantum fluctuations of the inflaton field during inflation are superseded
by its thermal fluctuations, due to a subdominant thermal bath, generated
and maintained by the dissipative effects. At the end of inflation, this thermal
bath suffices to reheat the Universe, thereby overcoming one of the major
problems of quintessential inflation model-building. Indeed, reheating cannot be
due to inflaton decay, as in conventional inflation, because the inflaton must
survive until today. A number of reheating mechanisms have been put forward, the
most important of which are gravitational reheating \cite{gravreh},
instant preheating \cite{instant}, curvaton reheating \cite{curvreh} and
recently non-minimal reheating \cite{nonminreh} (also called Ricci reheating
\cite{ricci}). In most cases, an extra degree of freedom must be assumed,
which is coupled to the inflaton (instant reheating) or not (curvaton or
non-minimal reheating), the only exception being gravitational reheating, which
however is in danger of producing excessive tensors \cite{samigrav}. In this
paper, efficient reheating occurs naturally without any additional assumptions.

\section{Warm inflation}

The slow-roll equations in warm inflation are 
\begin{eqnarray}
  3H(1+Q)\dot\phi & \!\!\!\simeq \!\!\! & -V'\label{KGwarmSR}\\
{\rm and}\qquad
  \rho_r &  \!\!\!\simeq\!\!\! & \frac34 Q\dot\phi^2\,,
\label{warmSR}
\end{eqnarray}
where $H$ is the Hubble scale, $\rho_r$ is the density of the subdominant
radiation, \mbox{$Q\equiv\Upsilon/3H$} with $\Upsilon$ being the dissipation
coefficient and the dot (prime) denotes differentiation with respect to time
(the inflaton field).
The scalar power spectrum in warm inflation is \cite{warmPz}
\begin{equation}
{\cal P}_\zeta=\frac{H^2(1+Q)^2{\cal F}}{8\pi^2\varepsilon m_P^2}\,,
\label{Pz}
\end{equation}
where $\varepsilon$
is the inflationary slow-roll parameter (defined later, in Eq.~(\ref{SR}))
and
\begin{equation}
{\cal F}\equiv 1+2{\cal N}_*+\frac{T}{H}\frac{2\pi Q}{\sqrt{1+\frac{4\pi}{3}Q}}
\,,
\label{F}
\end{equation}
with \mbox{${\cal N}_*=(e^{H/T}-1)^{-1}$} being the statistical distribution of
the inflaton field at horizon crossing, and $T$ is the temperature of the
subdominant thermal bath during inflation.\footnote{There is a minor
  correction to $\cal F$ when $Q>0.1$ which we ignore.} In cold inflation,
\mbox{$Q,T=0$} and \mbox{${\cal F}=1$} so that Eq.~(\ref{Pz}) reduces to the
usual expression. However, in warm inflation \mbox{$T\gg H$} and so
\mbox{${\cal N}_*\simeq T/H\gg 1$}. As mentioned, we consider the weak
dissipative regime, where \mbox{$Q<1$}. In this case, Eq.~(\ref{F}) suggests
\mbox{${\cal F}\simeq 2(1+\pi Q)T/H$}. For the density of the subdominant
thermal bath we have
\begin{equation}
  \rho_r=\frac{\pi^2}{30}g_* T^4=\frac{\varepsilon QV}{2(1+Q)^2}\,,
\label{rhor}
\end{equation}
where $g_*$ is the effective relativistic degrees of freedom and we used
the slow-roll Friedman equation \mbox{$V\simeq 3m_P^2H^2$} and
Eqs.~(\ref{KGwarmSR}) and (\ref{warmSR}) in the last equation.
Combining Eqs.~(\ref{Pz}) and (\ref{rhor}) we arrive at 
\begin{equation}
  {\cal P}_\zeta=\frac{1}{4\pi^2}\left(\frac{45}{\pi^2g_*}\right)^{1/4}
\frac{Q^{1/4}(1+Q)^{3/2}(1+\pi Q)}{\varepsilon^{3/4}}
\left(\frac{H}{m_P}\right)^{3/2}.
\label{Pz+}
\end{equation}

Now, we consider the model at hand. Warm quartic chaotic inflation has recently
been studied in detail in Ref.~\cite{maretal} (for some other related works see
Ref.~\cite{other}). The only difference in our setup
is that there is a small gap between the inflation and the radiation era, during
which the Universe assumes an equation of state stiffer than radiation. However,
we find that this period is very brief and serves only to add about one efold in
$N_*$; the number of remaining efolds of inflation when the cosmological scales
exit the horizon. As a result, our findings follow closely the much more
elaborate Ref.~\cite{maretal}.

During inflation, Eq.~(\ref{V}) suggests \mbox{$V\simeq\lambda\phi^4$}. Then we
find
\begin{equation}
  \varepsilon\equiv\frac12 m_P^2\left(\frac{V'}{V}\right)^2
  =8\left(\frac{m_P}{\phi}\right)^2
  \quad{\rm and}\quad
  \eta\equiv m_P^2\frac{V''}{V}=12\left(\frac{m_P}{\phi}\right)^2=
  \frac32\,\varepsilon\,.
  \label{SR}
\end{equation}
The number of remaining efolds of inflation is
\begin{equation}
  N=\frac{1}{m_P^2}\int_{\phi_{\rm end}}^{\phi(N)}\frac{V(1+Q)}{V'}\,{\rm d}\phi
  \;\Rightarrow\; N=\frac{1+Q}{8m_P^2}\left(\phi^2(N)-\phi_{\rm end}^2\right)\,,
\label{N}
\end{equation}
where `end' denotes the end of inflation and we have taken that, during
slow-roll, \mbox{$Q\simeq\,$constant}. Warm inflation ends when
\mbox{$\varepsilon=1+Q$}, which gives
\begin{equation}
  \phi^2(N)=\frac{8(N+1)}{1+Q}m_P^2\,,
\label{phiN}
\end{equation}
with \mbox{$\phi_{\rm end}=\phi(N=0)<0$}.\footnote{Recall that, during inflation
\mbox{$\phi<0$} as it is clear from Eq.~(\ref{V}).}
Thus, we obtain
\begin{equation}
  \varepsilon=\frac{1+Q}{N+1}\,.
  \label{eps}
\end{equation}
Combining the above with Eq.~(\ref{Pz+}) we get
\begin{equation}
  {\cal P}_\zeta=\frac{1}{4\pi^2}\left(\frac{45}{\pi^2g_*}\right)^{1/4}
Q^{1/4}(1+Q)^{3/4}(1+\pi Q)(N_*+1)^{3/4}
\left(\frac{H}{m_P}\right)^{3/2},
\label{Pzfin}
\end{equation}
where $N_*$ is the remaining efolds of inflation when the cosmological scales
exit the horizon.
In addition, using that \mbox{$V=3m_P^2H^2=\lambda\,\phi^4(N)$} we find
\begin{equation}
\frac{H}{m_P}=\frac{8\sqrt\lambda}{\sqrt 3}\frac{N_*+1}{1+Q}\,.
\label{lambda}
\end{equation}
For the tensor-to-scalar ratio we obtain
\begin{equation}
  r\equiv\frac{{\cal P}_h}{{\cal P}_\zeta}=\frac{2}{\pi^2{\cal P}_\zeta}
  \left(\frac{H}{m_P}\right)^2\,,
\label{r}
\end{equation}
where \mbox{${\cal P}_h=\frac{2}{\pi^2}(H/m_P)^2$} is the tensor spectrum,
which is unaffected by dissipative effects. However, we should stress here that
considering warm inflation reduces the value of $r$ compared to cold inflation.
The reason is that, because \mbox{$T>H$} in warm inflation, the scalar
perturbations are due to thermal fluctuations of the inflaton field,
which dominate the field's quantum fluctuations. This means that, in warm
inflation the value of the scalar spectrum ${\cal P}_\zeta$ is enhanced compared
with cold inflation. Normalising ${\cal P}_\zeta$ with the observations
\mbox{${\cal P}_\zeta=2.10\times 10^{-9}$} \cite{planck} implies that we may
produce the observed curvature perturbation with a lower inflation scale,
meaning with a lower value of $H$. In turn, as shown in Eq.~(\ref{r}), this
corresponds to a lower value of $r$.

Finally, for the scalar spectral index, in the case of warm inflation
we have \cite{warmns}
\begin{equation}
  n_s-1=-\frac{17+9Q}{4(1+Q)^2}\;\varepsilon
  +\frac{3}{2(1+Q)}\;\eta-\frac{1+9Q}{4(1+Q)^2}\;\beta\,,
\label{nsQ}
\end{equation}
where \mbox{$\beta\equiv m_P^2\frac{\Upsilon'V'}{\Upsilon V}$}. Considering
that the dissipation coefficient does not depend on the inflaton
field \mbox{$\Upsilon\neq\Upsilon(\phi)$} (as in Ref.~\cite{maretal}) so that
\mbox{$\beta=0$} and using that \mbox{$\eta=\frac32\varepsilon$}
(cf. Eq.~(\ref{SR})) the above reduces to
\begin{equation}
n_s=1-\frac{2\varepsilon}{(1+Q)^2}=1-\frac{2}{(1+Q)(N_*+1)}\,,
  \label{ns}
\end{equation}  
where we also used Eq.~(\ref{eps}).

\section{End of inflation}

Now, let us focus at the end of inflation.
Using that at the end of inflation \mbox{$\varepsilon=1+Q$},
Eq.~(\ref{rhor}) readily gives
\begin{equation}
\rho_r^{\rm end}=\frac12\frac{Q}{1+Q}V_{\rm end}\;.
\label{rhorend}
\end{equation}
Using Eqs.~(\ref{warmSR})
and (\ref{rhorend}), the kinetic density of the inflaton field
at the end of inflation is
\begin{equation}
\rho_{\rm kin}^{\rm end}=\frac12\dot\phi_{\rm end}^2=\frac23\frac{\rho_r^{\rm end}}{Q}=
  \frac13\frac{V_{\rm end}}{1+Q}\,.
\label{rhokinend}
\end{equation}
Thus, the total density of the inflaton at the end of inflation is
\begin{equation}
\rho_\phi^{\rm end}=\rho_{\rm kin}^{\rm end}+V_{\rm end}=\frac{4+3Q}{3(1+Q)}V_{\rm end}\;.
\label{rhophiend}
  \end{equation}
From Eqs.~(\ref{rhorend}) and (\ref{rhophiend}) we find the density parameter
of radiation at the end of inflation
\begin{equation}
  \Omega_r^{\rm end}\equiv
  \left.\frac{\rho_r}{\rho}\right|_{\rm end}\simeq
  \left.\frac{\rho_r}{\rho_\phi}\right|_{\rm end}=
  \frac{3Q}{2(4+3Q)}\,,
\label{omega}
\end{equation}
where \mbox{$\rho=\rho_\phi+\rho_r$} and we considered
\mbox{$(\rho_r/\rho_\phi)_{\rm end}\ll 1$}.

Consider now, what happens after the end of inflation and until the
thermal bath generated due to dissipation, dominates the Universe and the
radiation era begins. For radiation we have \mbox{$\rho_r\propto a^{-4}$}, where
we considered that further dissipation is negligible and radiation is an
independent fluid. The same is true for the inflaton field itself, for which
\mbox{$\rho_\phi\propto a^{-3(1+w)}$}, where $w$ is its effective equation of
state, taken as constant for simplicity. Thus, the radiation density parameter
scales as \mbox{$\Omega_r=\rho_r/(\rho_r+\rho_\phi)\simeq
  \rho_r/\rho_\phi\propto a^{3w-1}$}, with \mbox{$\rho_r<\rho_\phi$}. Reheating
(denoted by `reh') is the moment when \mbox{$\rho_r=\rho_\phi$}, which means
\mbox{$\Omega_r^{\rm reh}=\frac12$}. Therefore, we find
\begin{eqnarray}
 & & \frac12\simeq\Omega_r^{\rm end}\left(\frac{a_{\rm reh}}{a_{\rm end}}\right)^{3w-1}
  \nonumber\\
\Rightarrow & &
\frac{T_{\rm reh}}{T_{\rm end}}=\frac{a_{\rm end}}{a_{\rm reh}}\simeq
\left(\frac{3Q}{4+3Q}\right)^{1/(3w-1)},
\label{TT}
\end{eqnarray}
where we used Eq.~(\ref{omega}) and that \mbox{$T\propto 1/a$}. Using that
\mbox{$\rho_r=\frac{\pi^2}{30}g_*T^4$} and Eq.~(\ref{rhorend}), the above gives
\begin{equation}
\frac{V_{\rm end}^{1/4}}{T_{\rm reh}}\simeq
\left(\frac{\pi^2g_*}{15}\right)^{1/4}\left(\frac{1+Q}{Q}\right)^{1/4}
\left(\frac{4+3Q}{3Q}\right)^{1/(3w-1)}.
\label{VT}
\end{equation}
When a period of stiff equation of state follows inflation, 
the value of $N_*$ obtains an addition, given by
\begin{equation}
\Delta N=\frac{3w-1}{3(1+w)}\ln\left(\frac{V_{\rm end}^{1/4}}{T_{\rm reh}}\right)\,,
\label{DN}
\end{equation}
where the ratio $V_{\rm end}^{1/4}/T_{\rm reh}$ is given by Eq.~(\ref{VT}) and $w$ is
the barotropic parameter of the Universe. As long as the radiation bath remains
subdominant, \mbox{$w=w_\phi$}, where $w_\phi$ is the barotropic parameter of the
inflaton field.

Let us obtain an estimate of how large $\Delta N$ is. To maximise the effect of
the period after inflation and before reheating, we make the approximation that
the field becomes kinetically dominated immediately after the end of inflation,
so that \mbox{$w_\phi=1$}. We consider the range
\begin{equation}
  0.001\leq Q<0.1\;.
  \label{Qrange}
\end{equation}
Then, taking also \mbox{$g_*=106.75$} which corresponds to
the standard model at high energies, Eqs.~(\ref{VT}) and (\ref{DN}) suggest
\mbox{$\Delta N\simeq 0.69 - 2.13$}. In Ref.~\cite{maretal} the number of
efolds that correspond to the cosmological scales was 58. Thus, in our case (we
have to add about one because of $\Delta N$) we find \mbox{$N_*+1\approx 60$}.

In the range shown in Eq.~(\ref{Qrange}) we also obtain the following.
Eq.~(\ref{Pzfin}) allows us to calculate $H$, using the fact that
\mbox{${\cal P}_\zeta=2.10\times 10^{-9}$} \cite{planck}. We find
\mbox{$H=(0.48-1.31)\times 10^{-5}m_P$}.
Using these values in Eq.~(\ref{lambda}) we obtain
\mbox{$\lambda=(0.37-2.24)\times 10^{-15}$},
which is close to the results found in Ref.~\cite{maretal}. For the
inflationary observables we find the following. 
Eq.~(\ref{ns}) suggests \mbox{$n_s=0.967-0.969$} which is excellent (it
falls within the 1-$\sigma$ contours of the Planck observations \cite{planck}),
while Eq.~(\ref{r}) gives
\mbox{$r=0.0023-0.0166$},
which is potentially observable in the near future and 
satisfies the observational constrain \mbox{$r<0.07$}~\cite{planck}.

\section{Quintessence}

After inflation the field runs down the potential until it freezes.\footnote{%
  After inflation, the field transverses a distance of several Planck-scales in
  field space. Because of this we expect the dissipation processes to differ
  substantially compared to the period of inflation. This is why we can assume
  that dissipation is suppressed away from the inflation slope and is negligible
  afterwards.}
This occurs
even if the field is subdominant to radiation, so it does not matter that much
that the field remains dominant after inflation only for about an efold or two.
As we mentioned before, the field is kinetically dominated until it freezes. In
this case, it has been shown in Ref.~\cite{QIalpha} that the value where the
field freezes is solely determined by the density parameter of radiation at the
end of inflation and it is given by
\begin{equation}
\phi_F=\phi_{\rm end}+\sqrt{\frac23}\left(1-\frac32\ln\Omega_r^{\rm end}\right)m_P\;.
\label{phiF}
\end{equation}
Using Eqs.~(\ref{phiN}) and (\ref{omega}) the above can be recast as
\begin{equation}
  \phi_F=\left[-\frac{2\sqrt 2}{\sqrt{1+Q}}+\sqrt\frac23+\sqrt\frac32
\ln\left(\frac{2(4+3Q)}{3Q}\right)\right]m_P\;.
\label{phiFQ}
\end{equation}  
In the range shown in Eq.~(\ref{Qrange}) we find
\mbox{$\phi_F=(2.23-7.65)\,m_P$}.
Since \mbox{$\phi_F\gg M$} we are deep
down the quintessential tail of the potential. So we have
\mbox{$V\simeq\lambda M^8/\phi^4$} and the field now acts as IPL quintessence.

If quintessence remained frozen until the present, its residual potential
density would act as an effective cosmological constant. If that were the case,
then the value of this residual potential density must be such in order to
explain the dark energy observations. In turn, this requirement would allow the
calculation of the value of $M$. Indeed, assuming that quintessence remains
frozen we should demand that
\begin{equation}
  V(\phi_F)=\frac{\lambda M^8}{\phi_F^4}=\Omega_\Lambda\rho_0
  \simeq(2.25\times 10^{-3}\,{\rm eV})^4\,,
\label{VF}
\end{equation}
where \mbox{$\Omega_\Lambda\simeq 0.692$} \cite{planck} is the density parameter
of dark energy at present and \mbox{$\rho_0=0.864\times 10^{-29}
  \frac{\rm g}{\rm cm^3}$} \mbox{$=3.72\times 10^{-47}\,$GeV$^4$} is the
current density of the Universe. Using the values we have obtained, namely
\mbox{$\phi_F=(2.23-7.65)\,m_P$} and
\mbox{$\lambda=(0.37-2.24)\times 10^{-15}$},
the above suggests
\mbox{$M=(2.96-4.38)\times 10^5\,$GeV},
which is a rather reasonable intermediate energy scale. 

However, our model is thawing quintessence \cite{thawing}, which means that
there is an attractor solution, which the field unfreezes and tries to follow,
when its density \mbox{$\rho_F=V(\phi_F)$} becomes comparable to the attractor
density, meaning the density that the field would have were it
following the attractor. For IPL quintessence the attractor is called a tracker
and it is an exact solution of the Klein-Gordon equation. For a quartic IPL
quintessence of the form \mbox{$V=\hat M^8/\phi^4$}, the tracker solution is
\cite{tracker}
\begin{equation}
  \phi_A=\left(3\hat M^4 t\right)^{1/3}.
\label{tracker}
\end{equation}
This solution assumes a matter dominated Universe and is valid only when
quintessence is subdominant. In the range shown in Eq.~(\ref{Qrange}), we have
\mbox{$\hat M=\lambda^{1/8}M=(3.49-6.46)\,$TeV}.

As a zeroth-order approximation we consider that the quintessence field remains
frozen provided its density \mbox{$\rho_A>\rho_F=V(\phi_F)$} at present.
This requirement provides a lower bound on the value of $\phi_F$.
Indeed, using Eq.~(\ref{tracker}), we find
\begin{equation}
  \rho_A=\frac12\dot\phi_A^2+V(\phi_A)
  =\frac32\left(\frac{\hat M}{3\,t}\right)^{4/3}=
  \frac32 V(\phi_A)\,.
\label{rhoA}
\end{equation}
Evaluating the above at the present time $t_0$ we find
\begin{equation}
  \phi_F>(2/3)^{1/4}\phi_A(t_0)=(2/3)^{1/4}(3\hat M^4 t_0)^{1/3}\,.
\label{phiFbound}
\end{equation}
Using our findings, namely that
\mbox{$\hat M=(3.49-6.46)\,$TeV}
and that \mbox{$t_0=13.8\,$Gy$\,=6.62\times 10^{41}\,$GeV$^{-1}$} we obtain
\mbox{$\phi_A(t_0)=(2.74-6.22)m_P$}, which results in the bound
\mbox{$\phi_F>(2.48-5.62)\,m_P$}.
This is very close to the values we have found
\mbox{$\phi_F=(2.23-7.65)\,m_P$}.
The ratio of the corresponding densities today is
\begin{equation}
  \frac{\rho_A(t_0)}{V(\phi_F)}=
  0.66-3.44\,.
\label{Vratio}
\end{equation}

\begin{figure}

\begin{center}
\vspace{-4cm}

\leavevmode
\hbox{%
\epsfxsize=5.5in
\epsffile{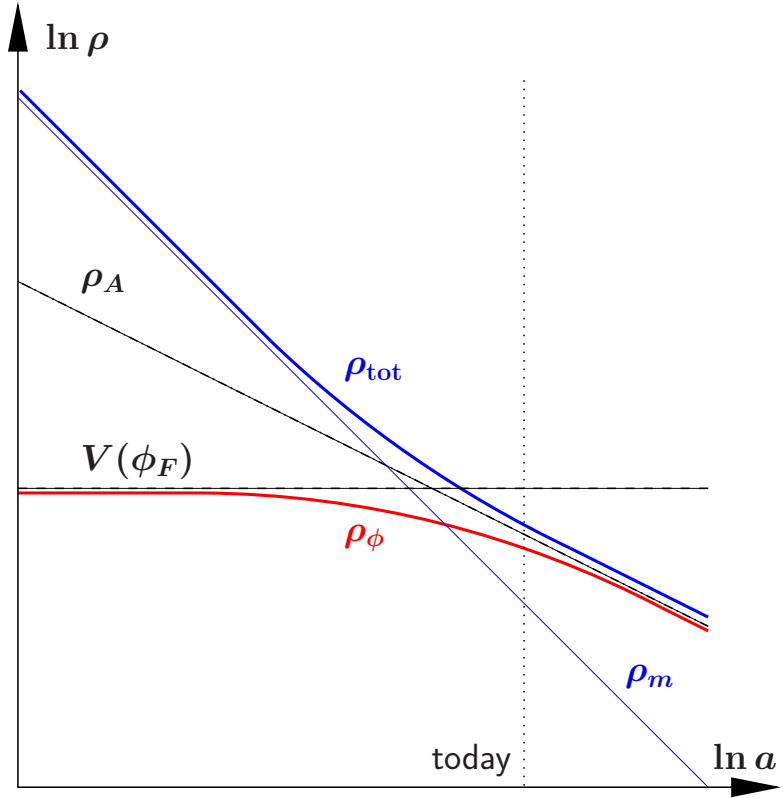}}
\vspace{-5cm}
\caption{
  Schematic log-log plot of the evolution of densities in thawing quintessence
with \mbox{$V(\phi)\propto\phi^{-4}$}. 
\mbox{$V(\phi_F)=\,$}constant is depicted with the horizontal dashed
line. The attractor (tracker) \mbox{$\rho_A\propto a^{-2}$} is
depicted with the slanted dot-dashed line. The slanted thin solid line (blue)
depicts the density of matter \mbox{$\rho_m\propto a^{-3}$}, while the lower
thick solid line (red) depicts $\rho_\phi$ and the upper thick solid line (blue)
depicts the total density \mbox{$\rho_{\rm tot}=\rho_m+\rho_\phi$}. The present
time is shown with the vertical dotted line. As evident in the figure, recently
the density of quintessence unfreezes in an attempt to follow the tracker.
Today \mbox{$\rho_m<\rho_\phi<\rho_A<V(\phi_F)$}. Note however, that the tracker
solution is not valid after the end of the matter era and quintessence is
expected to undergo slow-roll down its potential.}
\label{wqifig}
\end{center}
\end{figure}


However, the actual situation is more complicated. Indeed, when
\mbox{$V(\phi_F)\simeq\rho_A$},
we expect quintessence to unfreeze and start slow-rolling in an attempt to
follow the tracker, as shown in Fig.~\ref{wqifig}. This however, is undermined
by the fact that the tracker solution is losing its validity at present because
we are no more in the pure matter era and the dark energy is about to dominate
the Universe. Therefore, we should numerically investigate the problem, which
may need a slightly modified value of $M$ to work.

\begin{figure}

\begin{center}

\leavevmode
\hbox{%
\epsfxsize=5.5in
\epsffile{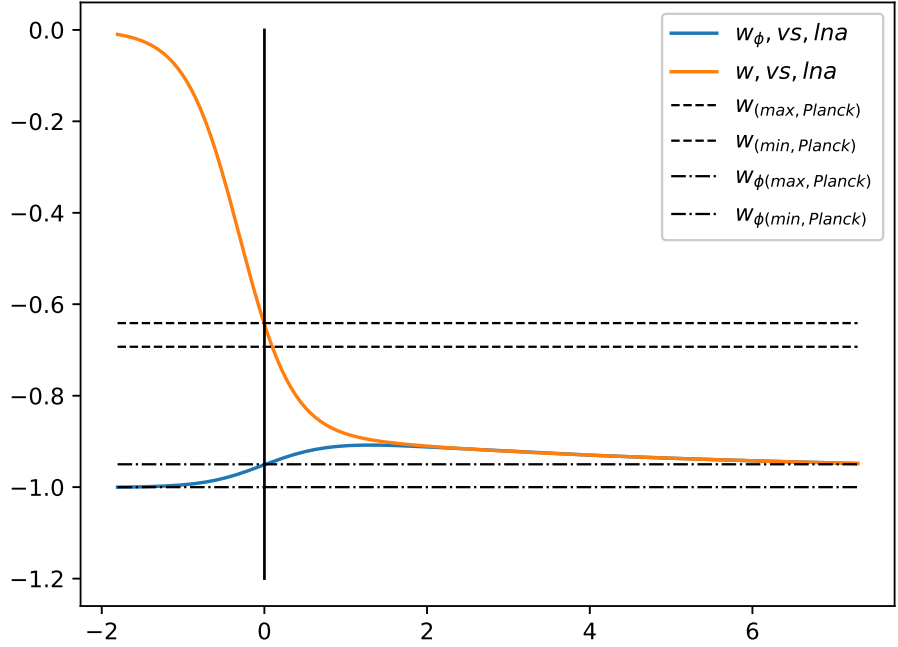}}
\caption{
  Behaviour of the barotropic parameter of quintessence $w_\phi$ (lower solid
  curve -blue) and of the whole Universe $w$ (upper solid curve - orange)
  as a function of the logarithm of the scale factor $\ln a$, which
  is normalised to unity today \mbox{$a_0=1$}. We see that
  originally the Universe is in the matter era with \mbox{$w=0$} and the
  quintessence field is frozen with constant density $V(\phi_F)$, such that
  \mbox{$w_\phi=-1$}. However, when approaching the present time (depicted by the
  vertical solid line - black) the quintessence unfreezes and
  \mbox{$w_\phi(t)>-1$}, while it also begins to dominate the Universe so that
  \mbox{$w(t)<0$}. Choosing the limiting case \mbox{$\phi_F=6.80\,m_P$} the
  present values of $w_\phi$ and $w$ satisfy the Planck bounds, depicted by the
  horizontal lines. In the future, quintessence becomes fully dominant so
  \mbox{$w\approx w_\phi$}, while it slow-rolls down the quintessential tail of
  the scalar potential, ever more slowly, approximating
  \mbox{$w=w_\phi\rightarrow -1$}. It is clear that both $w_\phi$ and $w$ are
  running at present, with \mbox{$w_a\equiv-dw_\phi/da|_{a=a_0}<0$}.}
\label{leonora}
\end{center}
\end{figure}

Preliminary study is optimistic and the
resulting barotropic parameter for dark energy is within the observational
bounds \mbox{$-1\leq w_\phi\leq -0.95$} \cite{planck}.\footnote{If quintessence
  were following the tracker solution in Eq.~(\ref{tracker}), then we would have
  \mbox{$\rho_\phi\propto V\propto\phi^{-4}\propto t^{-4/3}\propto a^{-2}$}, which
  would imply a barotropic parameter \mbox{$w_\phi=-1/3$}, that is unacceptable.}
The same is true of its running. In fact, the scenario presents some distinct
observational signatures, because a potentially varying $w_\phi$ is to be probed
by forthcoming observations, such as EUCLID. We find that
\mbox{$\phi_F\geq 6.80\,m_P$} and \mbox{$0>w_a\geq -0.0659$}, where
\mbox{$w_a\equiv-dw_\phi/da|_{a=a_0}$}, (which is well within the Planck
bounds \mbox{$w_a=-0.28^{+0.31}_{-0.27}$} \cite{planck}), with
\mbox{$\hat M=6.25\,$TeV} and \mbox{$a_0\equiv a(t_0)$} being the scale factor
at the present time. The behaviour of the barotropic parameter of quintessence
$w_\phi$ and of the Universe $w$ is shown in Fig.~\ref{leonora} for the limiting
case \mbox{$\phi_F=6.80\,m_P$} (where \mbox{$w_a=-0.0659$}). We see that the
values found satisfy the Planck bounds. 

From Eq.~(\ref{phiFQ}), taking \mbox{$\phi_F=6.80\,m_P$} corresponds to
choosing \mbox{$Q=0.002$}. Then, Eq.~(\ref{Pzfin}) gives
\mbox{$H=1.16\times 10^{-5}\,m_P$}. Using this, Eq.~(\ref{lambda}) suggests
\mbox{$\lambda=1.77\times 10^{-15}$}. For the inflationary observables,
Eq.~(\ref{ns}) results in \mbox{$n_s=0.967$} and Eq.~(\ref{r}) gives
\mbox{$r=0.0130$}. Both comfortably satisfy the observational bounds.
The value \mbox{$\hat M=6.25\,$TeV} suggests that
\mbox{$M=\lambda^{-1/8}\hat M=4.36\times 10^5\,$GeV}. Finally, the potential
density when the field is still frozen~is
\begin{equation}
  V(\phi_F)=\frac{\hat M^8}{\phi_F^4}=(2.36\times 10^{-3}\,{\rm eV})^4\,.
\label{VFfin}
\end{equation}
Comparing the above with $\Omega_\Lambda\rho_0$ as given in Eq.~(\ref{VF}) we have
\mbox{$V(\phi_F)/\Omega_\Lambda\rho_0=(\frac{2.36}{2.25})^4=1.21>1$}, which agrees
with the expectation that the field has unfrozen and its density at present is
smaller than $V(\phi_F)$, as suggested by Fig.~\ref{wqifig}.

Before concluding, we briefly discuss the dissipative coefficient.
By considering \mbox{$\Upsilon\neq\Upsilon(\phi)$} we implicitly considered
the case when \mbox{$\Upsilon=C_TT$}, as in Ref.~\cite{maretal} (see also
Ref.~\cite{rosa}).
Then we find
\begin{equation}
  C_T=3QH/T\,.
\label{CT}
\end{equation}
In order to have warm inflation \mbox{$T>H$}. Indeed, in Ref.~\cite{maretal}
it is found that \mbox{$T/H={\cal O}(10)$}. Thus, with \mbox{$Q=0.002$},
Eq.~(\ref{CT}) suggests \mbox{$C_T\sim 10^{-3}$}.

\section{Conclusions}

In this paper we have discussed warm quintessential inflation. As a toy model
we have considered the original quintessential inflation model of
Ref.~\cite{PV}, which is shown in Eq.~(\ref{V}). We stress however, that the
scalar potential in Eq.~(\ref{V}) is only experienced during the inflation and
quintessence regimes when \mbox{$|\phi|\sim m_P$}, while the field is
kinetically dominated when \mbox{$|\phi|\ll m_P$}, which means that it is
oblivious of the potential, when crossing the origin. Because of this fact,
the exact form of the potential in Eq.~(\ref{V}) when \mbox{$|\phi|\ll m_P$}
should not be taken too seriously. In fact, warm quintessential inflation
could in principle be a possibility when considering other models of
quintessential inflation in the literature (see for example Ref.~\cite{QIalpha}
and references therein).

The warm quintessential inflation model presented here appears promising
for a more thorough investigation, especially of the time near the end of
inflation and until reheating (which determines $N_*$ and indirectly affects the
inflationary observables $n_s$ and~$r$) and also of the time near the present,
where there is connection with the dark energy observations. It is our
intention to pursue this study, but we thought that the basic idea should be
put out there first. Our promising findings suggest that modelling warm
quintessential inflation can be a fruitful new avenue, especially when
attempting to reconcile inflation, dark energy and the swampland conjectures.

Our paper appeared first but it was soon followed by Ref.~\cite{rosa2}, which
studies a very similar model. There are aspects of the system studied where each
paper focuses more than the other (for example, our work is more elaborate
regarding the behaviour of the quintessence field at present) and, in that
sense, both works complement each other.

\paragraph{Acknowledgements}
We would like to thank Vahid Kamali and Charlotte Owen for discussions. KD was
supported (in part) by the Lancaster-Manchester-Sheffield Consortium for
Fundamental Physics under STFC grant: ST/L000520/1.

\end{document}